\documentclass[seceq,preprint]{ptptex}

\usepackage{amsmath}
\usepackage{amssymb}
\usepackage[dvipdfmx]{graphicx}

\preprintnumber{IU-TH-13}

\title{
Reality from maximizing overlap 
in the future-included real action theory
}

\author{%
Keiichi \textsc{Nagao}\footnote{E-mail: keiichi.nagao.phys@vc.ibaraki.ac.jp}
and Holger Bech \textsc{Nielsen}\footnote{E-mail: hbech@nbi.dk}
}

\inst{
$^{*)}$Faculty of Education, Ibaraki University, Mito 310-8512 
Japan \\
$^{**)}$Niels Bohr Institute, University of Copenhagen, 
Blegdamsvej 17, 2100 Copenhagen $\O$, Denmark
}

\abst{%
In the future-included real action theory whose path runs 
over not only past but also future, we demonstrate a theorem, which states that  
the normalized matrix element of a Hermitian operator $\hat{\cal O}$ 
defined in terms of the future state at the final time $T_B$ and 
the fixed past state at the initial time $T_A$ 
becomes real for the future state selected such that 
the absolute value of the transition amplitude from 
the past state to the future state is maximized. 
This is a special version of our previously proposed theorem for 
the future-included complex action theory. 
We find that though the maximization principle 
leads to the reality of the normalized matrix element in the future-included real action 
theory, it does not specify the future and past states so much as in the case of the 
future-included complex action theory. 
In addition, we argue that 
the normalized matrix element seems to be more natural 
than the usual expectation value. 
Thus we speculate that the functional integral formalism of quantum theory 
could be most elegant in the future-included complex action theory. 
}

\begin{document}

\maketitle

\section{Introduction}

Quantum theory is usually formulated so that in the path integral 
the time integration is performed over the period between the initial time $T_A$ and 
some specific time, say, the present time $t$. 
Let us call this usual formulation the future-not-included theory. 
For quantum theory we could consider another formulation, the future-included theory, 
in which not only the past state $| A(T_A) \rangle$ at the initial time $T_A$ 
but also the future state $| B(T_B) \rangle$ at the final time $T_B$ is given at first, 
and the time integration is performed over the whole period 
from the past to the future. 
In the future-included theory, 
the normalized matrix element~\cite{Bled2006}\footnote{$\langle \hat{\cal O} \rangle^{BA}$ is called the weak value~\cite{AAV} in the context of the real action theory (RAT), and it has been intensively studied. 
For details, see Ref.~\cite{review_wv} and references therein.} 
$\langle \hat{\cal O} \rangle^{BA} 
\equiv \frac{ \langle B(t) |  \hat{\cal O}  | A(t) \rangle }{ \langle B(t) | A(t) \rangle }$, 
where $t$ is an arbitrary time ($T_A \leq t \leq T_B$), 
seems to have a role of an expectation value of an operator $\hat{\cal O}$. 
Indeed, in Refs.~\cite{Nagao:2012mj,Nagao:2012ye} 
we argued in the case of the action being complex that,  
if we regard $\langle \hat{\cal O} \rangle^{BA}$ 
as an expectation value in the future-included theory, 
we obtain the Heisenberg equation, Ehrenfest's theorem, 
and a conserved probability current density. 
The complex action theory (CAT) is an attempt to describe quantum theory 
so that the action is complex at a fundamental level, but effectively looks real~\cite{Bled2006}. 
It has been investigated intensively 
with the expectation that the imaginary part of the action 
would give some falsifiable predictions~\cite{Bled2006,Nielsen:2007ak,Nielsen:2008cm,Nielsen:2005ub,Nielsen:2007mj,newer1,Vaxjo2009,newer2,Nielsen2010qq,degenerate,Nielsen2009hq,Bled2010B,Nagao:2010xu,Nagao:2011za, Nagao:2011is,Nagao:2013eda, Nagao:2012mj,Nagao:2012ye, Nagao:2015bya}.\footnote{The corresponding Hamiltonian $\hat{H}$ is generically non-normal, 
not restricted to the class of PT-symmetric non-Hermitian Hamiltonians, as studied in Refs.~\cite{Bender:1998ke,Bender:1998gh,Mostafazadeh_CPT_ip_2002,Mostafazadeh_CPT_ip_2003}.}


In the CAT the imaginary parts of the eigenvalues 
of $\hat{H}$ are supposed to be bounded from above 
to avoid the Feynman path integral
$\int e^{\frac{i}{\hbar}S} {\cal D} \text{path}$ 
being divergently meaningless. 
In Ref.~\cite{Nagao:2015bya} 
we proposed a theorem that states that, 
provided that an operator $\hat{\cal O}$ is $Q$-Hermitian, i.e., 
Hermitian with regard to the proper inner 
product $I_Q$ which makes the given Hamiltonian normal 
by using an appropriately chosen Hermitian operator $Q$, 
the normalized matrix element defined with $I_Q$ becomes real and 
time-develops under a $Q$-Hermitian Hamiltonian for the past and future states selected such that the absolute value of the transition amplitude defined with $I_Q$ from the past state to the future state is maximized.
We call this way of thinking the maximization principle. 
In Ref.~\cite{Nagao:2015bya} we gave the proof of the theorem in the case of non-normal 
Hamiltonians $\hat{H}$  
by finding that 
essentially only terms associated with the largest imaginary parts 
of the eigenvalues of the Hamiltonian contribute most 
to the absolute value of the transition amplitude defined with $I_Q$, 
and that the normalized matrix element defined with $I_Q$ 
for such maximizing states becomes an expression similar to 
an expectation value defined with $I_Q$ in the future-not-included theory.

In this letter we study the above theorem in the case of Hermitian Hamiltonians, i.e., 
in the future-included real action theory (RAT). 
After briefly reviewing the theorem in the future-included CAT, 
we show that even in the future-included RAT, 
being supplemented by the maximization principle, 
the normalized matrix element 
$\langle \hat{\cal O} \rangle^{BA}$ becomes real, though it is generically complex by definition. 
This study strongly supports using $\langle \hat{\cal O} \rangle^{BA}$ instead of 
$\langle \hat{\cal O} \rangle^{AA} 
\equiv \frac{ \langle A(t) |  \hat{\cal O}  | A(t) \rangle }{ \langle A(t) | A(t) \rangle }$ 
for an expectation value in the future-included RAT, 
and supplements the proof of our previously proposed theorem~\cite{Nagao:2015bya} 
in this special case of Hamiltonians being Hermitian. 
In addition, we find that though the maximization principle leads to the reality 
of the normalized matrix element in the future-included RAT, it does not specify 
the future and past states so much as in the case of the future-included CAT. 
Furthermore, arguing that 
the normalized matrix element $\langle \hat{\cal O} \rangle^{BA}$ seems to be more natural 
than the usual expectation value $\langle \hat{\cal O} \rangle^{AA}$, 
we speculate that the functional integral formalism of quantum theory 
could be most elegant in the future-included CAT.

\section{Review of the maximization principle in the future-included complex action theory}

In the case of non-normal Hamiltonians, 
since the eigenstates of $\hat{H}$, $| \lambda_i \rangle (i=1,2,\dots)$ 
obeying $\hat{H} | \lambda_i \rangle = \lambda_i | \lambda_i \rangle$,  are not orthogonal to 
each other in the usual inner product $I$, 
we introduce the proper inner product $I_Q$~\cite{Nagao:2010xu,Nagao:2011za} 
that makes $\hat{H}$ normal with respect to it.\footnote{Similar inner products are also studied 
in Refs.~\cite{Geyer,Mostafazadeh_CPT_ip_2002,Mostafazadeh_CPT_ip_2003}.}  
Then $| \lambda_i \rangle (i=1,2,\dots)$ 
become orthogonal to each other with regard to $I_Q$, which is defined 
for arbitrary kets $|u \rangle$ and $|v \rangle$ as 
$I_Q(|u \rangle , |v \rangle) \equiv \langle u |_Q v \rangle 
\equiv \langle u | Q | v \rangle$, 
where $Q$ is a Hermitian operator obeying $\langle \lambda_i  |_Q \lambda_j \rangle = \delta_{ij}$. 
We choose $Q=(P^\dag)^{-1} P^{-1}$ by using 
the diagonalizing operator 
$P=(| \lambda_1 \rangle , | \lambda_2 \rangle , \ldots)$, by which  
$\hat{H}$ is diagonalized as 
$\hat{H} = PD P^{-1}$, where $D=\text{diag}(\lambda_1, \lambda_2, \dots)$. 
In addition, we define the $Q$-Hermitian conjugate $\dag^Q$ of 
an operator $A$ by $\langle \psi_2 |_Q A | \psi_1 \rangle^* \equiv \langle \psi_1 |_Q A^{\dag^Q} | \psi_2 \rangle$, so $A^{\dag^Q} \equiv Q^{-1} A^\dag Q$. 
We introduce $\dag^Q$ for kets and bras as 
$| \lambda \rangle^{\dag^Q} \equiv \langle \lambda |_Q $ and 
$\left(\langle \lambda |_Q \right)^{\dag^Q} \equiv | \lambda \rangle$, by which we can 
manipulate $\dag^Q$ in the same way as the usual $\dag$.
If $A$ obeys $A^{\dag^Q} = A$, $A$ is $Q$-Hermitian. 
Since 
$P^{-1}=
\left(
 \begin{array}{c}
      \langle \lambda_1 |_Q     \\
      \langle \lambda_2 |_Q     \\
      \vdots 
 \end{array}
\right)$ 
satisfies $P^{-1} \hat{H} P = D$ and 
$P^{-1} \hat{H}^{\dag^Q} P = D^{\dag}$, 
$\hat{H}$ is $Q$-normal, 
$[\hat{H}, \hat{H}^{\dag^Q} ] = P [D, D^\dag ] P^{-1} =0$.  
Thus the inner product $I_Q$ makes $\hat{H}$ 
$Q$-normal. 
We can decompose $\hat{H}$ as 
$\hat{H}=\hat{H}_{Qh} + \hat{H}_{Qa}$, 
where $\hat{H}_{Qh}= \frac{\hat{H} + \hat{H}^{\dag^Q} }{2}$ and 
$\hat{H}_{Qa} = \frac{\hat{H} - \hat{H}^{\dag^Q} }{2}$ are 
$Q$-Hermitian and anti-$Q$-Hermitian parts of $\hat{H}$, respectively.


In Ref.~\cite{Nagao:2015bya}, we adopted the proper inner product $I_Q$ 
for all quantities, and investigated the future-included 
theory~\cite{Bled2006,Nagao:2012mj,Nagao:2012ye}. 
It is described by using 
the future state $| B (T_B) \rangle$ at the final time $T_B$ 
and the past state $| A (T_A) \rangle$ at the initial time $T_A$, 
where $| A (T_A) \rangle$ and $| B (T_B) \rangle$ are supposed to 
time-develop as follows: 
\begin{eqnarray}
&&i \hbar \frac{d}{dt} | A(t) \rangle = \hat{H} | A(t) \rangle , \label{schro_eq_Astate} \\
&&
i \hbar \frac{d}{dt} | B(t) \rangle = {\hat{H}}^{\dag^Q} | B(t) \rangle 
\quad \Leftrightarrow \quad 
-i \hbar \frac{d}{dt} \langle B(t) |_Q  
= \langle B(t) |_Q  \hat{H} , 
\label{schro_eq_Bstate} 
\end{eqnarray}
and the normalized matrix element is given by 
\begin{equation}
\langle \hat{\cal O} \rangle_Q^{BA} 
\equiv \frac{ \langle B(t) |_Q  \hat{\cal O}  | A(t) \rangle }{ \langle B(t) |_Q A(t) \rangle }. 
\end{equation} 
If we change the notation of $\langle B(t) |$ such that it absorbs $Q$, it can be expressed simply as 
$\langle \hat{\cal O} \rangle^{BA}$~\cite{Bled2006}. 
This is called the weak value~\cite{AAV, review_wv} in the RAT, where $Q=1$. 
This quantity is a strong candidate for an expectation value 
in the future-included CAT, 
because if we regard $\langle \hat{\cal O} \rangle^{BA}$ 
as an expectation value in the future-included CAT, 
then we obtain the Heisenberg equation, Ehrenfest's theorem, 
and a conserved probability current density~\cite{Nagao:2012mj,Nagao:2012ye}.


Utilizing the proper inner product $I_Q$, 
we proposed the following theorem in Ref.~\cite{Nagao:2015bya} :

\vspace{0.5cm}
\noindent
{\bf Theorem 1.} 
{\em 
As a prerequisite, assume that a given Hamiltonian 
$\hat{H}$ is non-normal but diagonalizable 
and that the imaginary parts of the eigenvalues 
of $\hat{H}$ are bounded from above, 
and define a modified inner product $I_Q$ by means 
of a Hermitian operator $Q$ arranged so 
that $\hat{H}$ becomes normal with respect to $I_Q$. 
Let the two states $| A(t) \rangle$ and $ | B(t) \rangle$ 
time-develop according to the Schr\"{o}dinger equations 
with $\hat{H}$ and $\hat{H}^{\dag^Q}$ respectively: 
$|A (t) \rangle = e^{-\frac{i}{\hbar}\hat{H} (t-T_A) }| A(T_A) \rangle$, 
$|B (t) \rangle = e^{-\frac{i}{\hbar} {\hat{H}}^{\dag^Q} (t-T_B) } | B(T_B)\rangle$, 
and be normalized with $I_Q$ 
at the initial time $T_A$ and the final time $T_B$ respectively: 
$\langle A(T_A) |_{Q} A(T_A) \rangle = 1$,
$\langle B(T_B) |_{Q} B(T_B) \rangle = 1$. 
Next determine $|A(T_A) \rangle$ and $|B(T_B) \rangle$ so as to maximize 
the absolute value of the transition amplitude 
$|\langle B(t) |_Q A(t) \rangle|=|\langle B(T_B)|_Q \exp(-i\hat{H}(T_B-T_A)) |A(T_A) \rangle|$. 
Then, provided that an operator $\hat{\cal O}$ is $Q$-Hermitian, i.e., Hermitian 
with respect to the inner product $I_Q$, 
$\hat{\cal O}^{\dag^Q} = \hat{\cal O}$, 
the normalized matrix element of the operator $\hat{\cal O}$ defined by 
$\langle \hat{\cal O} \rangle_Q^{BA} 
\equiv
\frac{\langle B(t) |_Q \hat{\cal O} | A(t) \rangle}{\langle B(t) |_Q A(t) \rangle}$ 
becomes {\rm real} and time-develops under 
a {\rm $Q$-Hermitian} Hamiltonian. }


\vspace*{0.5cm}

\noindent
We call this way of thinking the maximization principle. 
The proof is given in Ref.~\cite{Nagao:2015bya}, but to see what the maximizing states are, 
let us expand $| A(t) \rangle$ and $| B(t) \rangle$ in terms of the eigenstates $| \lambda_i \rangle$ 
as follows: 
\begin{eqnarray}
&&|A (t) \rangle = \sum_i a_i (t) | \lambda_i \rangle, \label{Aketexpansion}\\
&&|B (t) \rangle = \sum_i b_i (t) | \lambda_i \rangle,  \label{Bketexpansion}
\end{eqnarray}
where 
\begin{eqnarray}
&&a_i (t) = a_i (T_A) e^{-\frac{i}{\hbar}\lambda_i (t-T_A) }, \label{aitimedevelopment} \\ 
&&b_i (t) = b_i (T_B) e^{-\frac{i}{\hbar}\lambda_i^* (t-T_B) }. \label{bitimedevelopment} 
\end{eqnarray}
We express  $a_i(T_A)$ and $b_i(T_B)$ as 
$a_i(T_A)= | a_i(T_A) | e^{i \theta_{a_i}}$ and 
$b_i(T_B) = | b_i(T_B) | e^{i \theta_{b_i}}$, and introduce 
$T\equiv T_B - T_A$ and $\Theta_i \equiv \theta_{a_i} - \theta_{b_i} 
- \frac{1}{\hbar} T \text{Re} \lambda_i$. 
Since the imaginary parts of the eigenvalues 
of $\hat{H}$ are supposed to be bounded from above 
to avoid the Feynman path integral
$\int e^{\frac{i}{\hbar}S} {\cal D} \text{path}$ 
being divergently meaningless, we can imagine that some of $\text{Im} \lambda_i$ 
take the maximal value $B$, and denote 
the corresponding subset of $\{ i \}$ as $A$. 
Then, $| \langle B (t) |_Q A (t) \rangle |$ can take 
the maximal value $e^{\frac{1}{\hbar} B T}$ only under the following conditions: 
\begin{eqnarray}
&& \Theta_i  \equiv \Theta_c \quad \text{for $\forall i \in A$}, \label{max_cond_theta}  \\ 
&& \sum_{i \in A} | a_i (T_A) |^2 =\sum_{i \in A}|b_i (T_B)|^2 = 1,  \label{nc_ATABTB3} \\
&& |a_i (T_A)| = |b_i (T_B)|  \quad \text{for $\forall i \in A$}, \label{max_cond_ab} \\
&& | a_i (T_A) |  = | b_i (T_B) | =0 \quad \text{for $\forall i \notin A$} , \label{abinotinA0}  
\end{eqnarray}
and the states to maximize $| \langle B (t) |_Q A (t) \rangle |$, 
$| A(t) \rangle_{\rm{max}}$ and $| B(t) \rangle_{\rm{max}}$, are 
expressed as 
\begin{eqnarray}
&&|A (t) \rangle_{\rm{max}} = \sum_{i \in A} a_i (t) | \lambda_i \rangle , 
\label{Atketmax_sum_inA_ai} \\
&&|B (t) \rangle_{\rm{max}} = \sum_{i \in A} b_i (t) | \lambda_i \rangle, 
\label{Btketmax_sum_inA_bi} 
\end{eqnarray}
where $a_i (t)$ and $b_i (t)$ obey 
Eqs.(\ref{max_cond_theta})-(\ref{max_cond_ab}). 
In addition, introducing $| \tilde{A}(t) \rangle \equiv 
e^{-\frac{i}{\hbar}(t-T_A) \hat{H}_{Qh}} $ $| A(T_A) \rangle_{\rm{max}}$, 
which is normalized as 
$\langle \tilde{A}(t) |_Q \tilde{A}(t) \rangle = 1$ 
and obeys the Schr\"{o}dinger equation 
$i\hbar  \frac{d}{d t}| \tilde{A}(t) \rangle = \hat{H}_{Qh} | \tilde{A}(t) \rangle$, 
we find that the normalized matrix element 
$\langle \hat{\cal O} \rangle_Q^{BA}$  
for $| A(t) \rangle_{\rm{max}}$ and $| B(t) \rangle_{\rm{max}}$ is evaluated as 
$\langle \hat{\cal O} \rangle_Q^{BA} =
\langle \tilde{A}(t) |_Q \hat{\cal O} | \tilde{A}(t) \rangle 
\equiv 
\langle \hat{\cal O} \rangle_Q^{\tilde{A} \tilde{A}}$, 
which is real for $Q$-Hermitian $\hat{\cal O}$, and also 
that 
$\langle \hat{\cal O} \rangle_Q^{\tilde{A} \tilde{A}}$ time-develops 
under the $Q$-Hermitian Hamiltonian $\hat{H}_{Qh}$: 
$\frac{d}{dt} \langle \hat{\cal O} \rangle_Q^{\tilde{A} \tilde{A}} 
=\frac{i}{\hbar} \langle \left[ \hat{H}_{Qh}, \hat{\cal O} \right] \rangle_Q^{\tilde{A} \tilde{A}}$. 
Thus the maximization principle provides both 
the reality of $\langle \hat{\cal O} \rangle_Q^{BA}$ 
for $Q$-Hermitian $\hat{\cal O}$ and the $Q$-Hermitian Hamiltonian.


The theorem presented above is given for systems defined with such general Hamiltonians 
that they do not even have to be normal, so it can 
also be used for normal Hamiltonians 
in addition to non-normal Hamiltonians. 
For a normal Hamiltonian $\hat{H}$, $Q$ is the unit operator. 
So in such a case the above theorem becomes simpler with $Q=1$. 
There are two possibilities for such a case: 
one is that $\hat{H}$ is non-Hermitian but normal, and the other is that $\hat{H}$ is 
Hermitian. 
In both cases $Q=1$, but there is a significant difference between the two cases: 
in the former case there are imaginary parts of 
the eigenvalues of $\hat{H}$, but not in the latter case. 
Since the proof given in Ref.~\cite{Nagao:2015bya} depends on the existence of 
imaginary parts of the eigenvalues of $\hat{H}$,  
it is still valid in the former case, 
but not in the latter case. 
Therefore, to see that the theorem also holds in the case of Hermitian Hamiltonians, 
we need to investigate it separately. 


\section{Maximization principle in the future-included real action theory}

Now we study the theorem proposed in Ref.~\cite{Nagao:2015bya} 
in the future-included RAT, 
where the Hamiltonian is Hermitian, to elucidate what the theorem tells us 
in the case of the RAT. 
In this special case, $Q$ is the unit operator. 
We can consider three possibilities in the future-included RAT: 
One is that  $ | A(T_A) \rangle$ is given at first, and $ | B(T_B) \rangle$ is chosen 
by the maximization principle. Another is the reverse. The other is that both 
$| A(T_A) \rangle$ and $ | B(T_B) \rangle$ are partly given and chosen. 
However, we know empirically the second law of thermodynamics, and 
the future-included RAT is closer to the usual theory, i.e., 
the future-not-included RAT, than the future-included CAT. 
So in the future-included RAT we choose the first option. 
We suppose that $| A(t) \rangle$ is a given 
fixed state, and only $| B(t) \rangle$ is a random state, which should be chosen appropriately 
by the maximization principle, though in the future-included CAT 
both $| A(t) \rangle$ and $| B(t) \rangle$ are supposed to be random states at first. 
In the case of non-normal Hamiltonians, it is nontrivial to obtain the 
emerging $Q$-hermiticity for the Hamiltonian by the maximization principle, 
so this is indeed one of the points that we emphasize. 
On the other hand, in the case of Hermitian Hamiltonians, 
the hermiticity of the Hamiltonian is given at first, so it does not become the point that we claim.  
Taking into account these remarks, we write the theorem particular to the case of 
Hermitian Hamiltonians as follows: 

\vspace{0.5cm}
\noindent
{\bf Theorem 2.} 
{\em 
As a prerequisite, assume that a given Hamiltonian 
$\hat{H}$ is diagonalizable and Hermitian.  
Let the two states $| A(t) \rangle$ and $ | B(t) \rangle$ 
time-develop according to the Schr\"{o}dinger equation with $\hat{H}$: 
$|A (t) \rangle = e^{-\frac{i}{\hbar}\hat{H} (t-T_A) }| A(T_A) \rangle$, 
$|B (t) \rangle = e^{-\frac{i}{\hbar} {\hat{H}} (t-T_B) } | B(T_B)\rangle$, 
and be normalized at the initial time $T_A$ and the final time $T_B$ respectively: 
$\langle A(T_A) | A(T_A) \rangle = 1$, 
$\langle B(T_B) | B(T_B) \rangle = 1$. 
Next determine $|B(T_B) \rangle$ for the given $|A(T_A) \rangle$ so as to maximize 
the absolute value of the transition amplitude 
$|\langle B(t) | A(t) \rangle|=
|\langle B(T_B)| \exp(-\frac{i}{\hbar}\hat{H}(T_B-T_A)) | A(T_A) \rangle|$. 
Then, provided that an operator $\hat{\cal O}$
is Hermitian, $\hat{\cal O}^\dag = \hat{\cal O}$, 
the normalized matrix element of the operator $\hat{\cal O}$ defined by 
$\langle \hat{\cal O} \rangle^{BA} 
\equiv
\frac{\langle B(t) | \hat{\cal O} | A(t) \rangle}{\langle B(t) | A(t) \rangle}$ 
becomes {\rm real} and time-develops under the given Hermitian Hamiltonian. }

\vspace{0.5cm}

\noindent
This theorem can be proven more simply than the theorem for non-Hermitian Hamiltonians. 
Indeed, since  
the norms of $ |  A(t) \rangle$ and $ |  B(t) \rangle$ are constant in time 
in the case of Hermitian Hamiltonians,  
$\langle A(t) |  A(t) \rangle=\langle A(T_A) |  A(T_A) \rangle=1$, 
$\langle B(t) |  B(t) \rangle=\langle B(T_B) |  B(T_B) \rangle=1$, 
we can directly use an elementary property of linear space, 
and find that the final state to maximize $| \langle B (t) | A (t) \rangle |$, 
$| B(T_B) \rangle_{\rm{max}}$, 
is the same as $| A(t) \rangle$ up to a constant phase factor:   
\begin{eqnarray}
|B (t) \rangle_{\rm{max}} 
= 
e^{-i \Theta_c } | A(t) \rangle . \label{BmaxphaseA}
\end{eqnarray}
This phase factor presents the ambiguity of the maximizing state  $| B(t) \rangle_{\rm{max}}$, 
and shows that $| B(t) \rangle_{\rm{max}}$ is not determined uniquely. 
The normalized matrix element $\langle \hat{\cal O} \rangle^{BA}$ 
for the given $| A(t) \rangle$ and $| B(t) \rangle_{\rm{max}}$ becomes 
\begin{eqnarray}
\langle \hat{\cal O} \rangle^{B_{\rm{max}}A} 
&=&
\frac{ {}_{\rm{max}}\langle B(t) |  \hat{\cal O}  | A(t) \rangle }{ {}_{\rm{max}}\langle B(t) | A(t) \rangle } \nonumber \\
&=&
\langle A(t) |  \hat{\cal O}  | A(t) \rangle 
\equiv
\langle\hat{\cal O} \rangle^{AA} . 
\end{eqnarray} 
Thus we have seen that $\langle \hat{\cal O} \rangle^{BA}$ 
for the given $| A(t) \rangle$ and $| B(t) \rangle_{\rm{max}}$ 
has become the form of a usual average $\langle \hat{\cal O} \rangle^{AA}$, 
and so it becomes real for a Hermitian operator $\hat{\cal O}$. 
In addition, we see that 
$\langle \hat{\cal O} \rangle^{AA}$ time-develops 
under the Hermitian Hamiltonian $\hat{H}$ as 
$\frac{d}{dt} \langle \hat{\cal O} \rangle^{A A} 
=\frac{i}{\hbar} \langle \left[ \hat{H}, \hat{\cal O} \right]  \rangle^{A A}$. 
We can introduce the Heisenberg operator 
$\hat{\cal O}_{H}(t, T_A) \equiv e^{ \frac{i}{\hbar} \hat{H} (t-T_A) } 
\hat{\cal O} e^{ -\frac{i}{\hbar} \hat{H} (t-T_A)}$ 
by expressing $\langle \hat{\cal O} \rangle^{A A}$ as 
$\langle \hat{\cal O} \rangle^{AA}=  \langle A(T_A) | \hat{\cal O}_{H}(t, T_A)
 | A(T_A) \rangle$. 
This obeys the Heisenberg equation 
$i\hbar  \frac{d}{d t} \hat{\cal O}_{H}(t, T_A) 
= [ \hat{\cal O}_{H}(t, T_A) , \hat{H} ]$. 
We emphasize that the maximization principle 
provides the reality of $\langle \hat{\cal O} \rangle^{BA}$ for Hermitian $\hat{\cal O}$, 
though $\langle \hat{\cal O} \rangle^{BA}$ is generically complex by definition.

To see the differences from the case of non-Hermitian Hamiltonians more explicitly, 
let us expand $| A(t) \rangle$ and $| B(t) \rangle$ 
in the same way as Eqs.(\ref{Aketexpansion})-(\ref{bitimedevelopment}), 
and introduce $R_i \equiv |a_i (T_A)| |b_i (T_B)| \geq 0$. 
Then, since 
$\langle B (t) | A (t) \rangle = \sum_i R_i e^{i \Theta_i}$, 
$| \langle B (t) | A (t) \rangle |^2
= \sum_i R_i^2 + 2 \sum_{i<j} R_i R_j \cos(\Theta_i - \Theta_j)$ 
can take a maximal value only under the condition: 
\begin{equation}
\Theta_i  = \Theta_c 
\quad \text{for $\forall i $} \label{max_cond_thetareal} ,  
\end{equation}
and $| \langle B (t) | A (t) \rangle |^2 = \left( \sum_{i } R_i \right)^2  
\leq  \left\{ \sum_{i} \left( \frac{ |a_i (T_A)| + |b_i (T_B)|}{2} \right)^2 \right\}^2 =1$, 
where the second equality is realized for 
\begin{equation}
 |a_i (T_A)| = |b_i (T_B)|  \quad \text{for $\forall i $}. 
\label{max_cond_abreal}  
\end{equation}
In the last equality we have used this relation and 
the normalization conditions for $|A(T_A) \rangle$ and $|B(T_B) \rangle$: 
$\sum_{i } | a_i (T_A) |^2 = \sum_{i }|b_i (T_B)|^2 = 1$. 
Since the condition for maximizing $| \langle B (t) | A (t) \rangle |$ 
is represented by Eqs.(\ref{max_cond_thetareal}), (\ref{max_cond_abreal})\footnote{In the case of a system defined with non-Hermitian Hamiltonians, 
the condition for maximizing $| \langle B (t) |_Q A (t) \rangle |$ 
is represented by 
Eqs.(\ref{max_cond_theta})-(\ref{abinotinA0}),  
and 
essentially only the subset having the 
largest imaginary parts of the eigenvalues of $\hat{H}$ contributes most to 
the absolute value of the transition amplitude $| \langle B (t) |_Q A (t) \rangle |$~\cite{Nagao:2015bya}. 
This is quite in contrast to the present study of a system defined with Hermitian Hamiltonians, 
where the full set of the eigenstates of $\hat{H}$ can contribute to  $| \langle B (t) | A (t) \rangle |$.}, 
$| B(T_B) \rangle_{\rm{max}}$ is expressed as 
$|B (T_B) \rangle_{\rm{max}} = \sum_{i } b_i^{\rm{max}} (T_B) | \lambda_i \rangle$, 
where 
$b_i^{\rm{max}} (T_B) 
\equiv |a_i (T_A)|  e^{ i \left(  \theta_{a_i} - \frac{1}{\hbar} T \lambda_i - \Theta_c \right) }$. 
So $|B (t) \rangle_{\rm{max}} =\sum_{i } b_i^{\rm{max}} (t) | \lambda_i \rangle$, 
where $b_i^{\rm{max}} (t) = a_i (t) e^{-i \Theta_c }$, becomes Eq.(\ref{BmaxphaseA}).

The maximizing principle in the case of Hermitian Hamiltonians 
specifies all the diagonal 
in the space of  $| A(t) \rangle$ and $| B(t) \rangle$ up to the phase factor. 
It turns out to be ambiguous in the sense of there being no unique class of states chosen. 
On the other hand, in the case of non-Hermitian Hamiltonians, 
just one class of points of the diagonal is chosen by the maximization principle. 
We can maximize the absolute value of the transition amplitude 
$| \langle B (t) |_Q A(t) \rangle |$ even with regard to 
$| A(t) \rangle$, because the norm of $ |  A(t) \rangle$ is not constant. 
Though the norm is $1$ at the initial time $T_A$,  
it deviates a lot from $1$ in later time. 
For a very different $|| ~|  A(t) \rangle  ||$, we can have a very different 
$| \langle B (t) |_Q A(t) \rangle |$. Thus only one class of combinations comes to dominate it. 
Therefore, in the case of the RAT, 
the maximizing principle is not so predictive as in the case of the CAT. 
In the RAT we have more solutions, 
which naturally give us the freedom to have initial conditions.

\section{Discussion}

In the future-included RAT we studied 
the normalized matrix element for a Hermitian operator $\hat{\cal O}$, $\langle \hat{\cal O} \rangle^{BA}=\frac{\langle B(t) | \hat{\cal O} | A(t) \rangle}{\langle B(t) | A(t) \rangle}$, 
which is expected to be a strong candidate for an expectation value in the future-included theory, 
as we argued in the case of the CAT 
in Refs.~\cite{Bled2006,Nagao:2012mj,Nagao:2012ye}. 
We showed that, provided that $|A(t) \rangle $ and $|B(t) \rangle$ time-develop 
according to the Schr\"{o}dinger equation 
with a given Hermitian diagonalizable Hamiltonian $\hat{H}$ and 
are normalized at the initial time $T_A$ and at the final time $T_B$ respectively, 
$\langle \hat{\cal O} \rangle^{BA}$ becomes real 
for the given state $|A(t) \rangle $ and the chosen state $|B(t) \rangle^{\rm{max}}$ 
such that the absolute value of the transition amplitude 
$|\langle B(t)| A(t) \rangle|$ is maximized. 
If we formulate quantum mechanics by using the normalized matrix element 
$\langle \hat{\cal O} \rangle^{BA}$ in the future-included theory, 
it would be natural to take into account the maximization principle. 
This study supplements the proof of the theorem, which was previously proposed 
for a system defined with non-normal Hamiltonians in Ref.~\cite{Nagao:2015bya}, 
and supports the speculation that the normalized matrix element has a role of 
an expectation value in the future-included theory. 
We found that the specification of the future and past states by the maximization principle 
is more ambiguous in the RAT than in the CAT. 
This is due to the lack of imaginary parts of the eigenvalues of the Hamiltonian in the RAT.

It is only by using the maximization principle that we get $\langle \hat{\cal O} \rangle^{BA}$ real. 
If we do not use the maximization principle, then the normalized matrix element 
$\langle \hat{\cal O} \rangle^{BA}$ is generically complex by definition even in the RAT. 
Since $\langle \hat{\cal O} \rangle^{BA}$ is expected to have a role of an expectation value 
in the future-included theory, we would have a complex classical solution  
unless we use the maximization principle. 
This situation is analogous to the usual classical physics. 
To illustrate this analogy, let us consider a harmonic oscillator whose Lagrangian is written as 
$L=\frac{1}{2}m \dot{q}^2 - \frac{1}{2} m \omega^2 q^2$, where $m$ and $\omega$ are real. 
Its classical solution is given by 
$q=A e^{i\omega t} + B e^{-i\omega t}$, where $A$ and $B$ are complex coefficients, 
so this $q$ is complex.  
Imposing on it the initial condition $(q,\dot{q})|_{t=0} \in {\mathbf R}^2$ 
leads to $B=A^*$, and thus $q$ acquires reality. 
Even in more general models, equations of motion are differential equations, so 
solutions in classical physics are generically complex 
unless we put such initial conditions giving reality. 
Similarly, also in our theory, classical solutions become complex without the maximization principle. 
So the maximization principle could be regarded as a special type of initial (or final) condition. 
Indeed, it specifies the past and final states to some extent.

In the usual theory, i.e., the future-not-included RAT, 
$\langle \hat{\cal O} \rangle^{AA}$ is real for a Hermitian operator $\hat{\cal O}$ by definition. 
It is constructed by hand so that it is real. 
On the other hand, in our future-included theory, 
$\langle \hat{\cal O} \rangle^{BA}$ is not adjusted so, 
but it becomes real by our natural way of thinking: the maximization principle. 
In addition, $\langle \hat{\cal O} \rangle^{BA}$ is expressed 
more elegantly than $\langle \hat{\cal O} \rangle^{AA}$ in the functional integral form: 
\begin{equation}
\langle \hat{\cal O} \rangle^{BA} 
= \frac{\int {\cal D} \text{path}~ \psi_B^* \psi_A {\cal O} e^{ \frac{i}{\hbar} S[\text{path}] } }{\int {\cal D}  \text{path}~ \psi_B^* \psi_A e^{\frac{i}{\hbar} S[\text{path}]  }}. \label{funcint}
\end{equation}
In the future-not-included theory, $\langle \hat{\cal O} \rangle^{AA}$ does not have 
such a full functional integral expression for all time. 
So $\langle \hat{\cal O} \rangle^{BA}$ seems to be more natural than $\langle \hat{\cal O} \rangle^{AA}$. 
We have the possibility of presenting the functional integral 
by using $\langle \hat{\cal O} \rangle^{BA}$ with the maximization principle. 
This interpretation provides a more direct connection of functional integrals to measurable physics. 
There could be $\langle \hat{\cal O} \rangle^{BA}$ 
behind $\langle \hat{\cal O} \rangle^{AA}$ at the fundamental level, and 
$\langle \hat{\cal O} \rangle^{BA}$ could effectively look like $\langle \hat{\cal O} \rangle^{AA}$ 
by the maximization principle.

Thus we can speculate that the fundamental physics is given by 
the normalized matrix element $\langle \hat{\cal O} \rangle^{BA}$ supplemented by 
the maximization principle 
rather than by the usual expectation value $\langle \hat{\cal O} \rangle^{AA}$. 
The initial and final states are integrated over 
so that the biggest obtainable $|\langle B(t) | A(t) \rangle|$ 
comes to dominate. 
In the case of the RAT, 
one would obtain a superposition or mixture of 
all the pairs of $| A(t) \rangle$ and $| B(t) \rangle$ with $| B(t) \rangle = | A(t) \rangle$ 
up to a phase factor, so the maximization principle 
leaves the initial condition 
to be chosen arbitrarily. 
There is no natural way to have a unique solution; 
we need to consider some assumptions and put in the initial condition 
for $| A(T_A) \rangle$. 
This seems to be a little bit artificial. 
However, if we somewhat wildly assumed the existence of an imaginary part of the action, 
then we would naturally 
obtain a specific solution, i.e., a unique class of $| A(t) \rangle$ and $| B(t) \rangle$ 
by maximizing $|\langle B(t) |_Q A(t) \rangle|$, 
as studied in Ref.~\cite{Nagao:2015bya}. 
The CAT could tell what  $| A(T_A) \rangle$ should be in an elegant way via the maximization principle. 
In this sense, the future-included CAT seems to be nicer than the future-included RAT, 
though it still requires a bit of phenomenological adjustment of the imaginary part of the action to 
get a cosmologically or experimentally good initial condition for our universe.  
Therefore, we speculate that the functional integral formalism of quantum theory 
would be most elegant in the future-included CAT, which might look extreme but cannot be  excluded 
from a phenomenological point of view, as studied in Refs.~\cite{Nagao:2012mj, Nagao:2012ye}.  
Only the maximization principle would be needed in addition to the imaginary part of the action. 
In the CAT we would obtain a unique class of solutions and thus even the initial state 
$| A(T_A) \rangle$ would be chosen. This could be the unification of the initial condition and the whole dynamics.

\section*{Acknowledgements}

K.N. would like to thank the members and visitors of NBI 
for their kind hospitality. 
H.B.N. is grateful to NBI for allowing him to work at the institute as emeritus. 
In addition, they are both grateful to Yutaka Shikano, Philip Mannheim, and Izumi Tsutsui 
for useful discussions.


\end{document}